\documentclass{aastex62}

\usepackage{upgreek}
\usepackage{color}
\usepackage{verbatim}

\received{--}
\revised{--}
\accepted{--}

\submitjournal{ApJL}

\shorttitle{The magnetic field of the Pillars of Creation}
\shortauthors{Pattle et al.}

\begin{document}

\title{First observations of the magnetic field inside the Pillars of Creation: Results from the BISTRO survey}

\correspondingauthor{Kate Pattle}
\email{kpattle@gapp.nthu.edu.tw}

\author[0000-0002-8557-3582]{Kate Pattle}
\affil{Jeremiah Horrocks Institute, University of Central Lancashire, Preston PR1 2HE, United Kingdom}
\affil{Institute of Astronomy and Department of Physics, National Tsing Hua University, Hsinchu 30013, Taiwan}

\author{Derek Ward-Thompson}
\affil{Jeremiah Horrocks Institute, University of Central Lancashire, Preston PR1 2HE, United Kingdom}

\author{Tetsuo Hasegawa}
\affil{National Astronomical Observatory of Japan, 2-21-1 Osawa, Mitaka, Tokyo 181-8588, Japan}

\author{Pierre Bastien}
\affil{Centre de recherche en astrophysique du Qu\'ebec \& d\'epartement de physique, Universit\'e de Montr\'eal, C.P. 6128, Succ. Centre-ville, Montr\'eal QC, H3C 3J7, Canada}

\author{Woojin Kwon}
\affil{Korea Astronomy and Space Science Institute, 776 Daedeokdae-ro, Yuseong-gu, Daejeon 34055, Republic of Korea}
\affil{Korea University of Science and Technology, 217 Gajang-ro, Yuseong-gu, Daejeon 34113, Republic of Korea}

\author{Shih-Ping Lai}
\affil{Institute of Astronomy and Department of Physics, National Tsing Hua University, Hsinchu 30013, Taiwan}
\affil{Academia Sinica Institute of Astronomy and Astrophysics, P. O. Box 23-141, Taipei 10617, Taiwan}

\author{Keping Qiu}
\affil{School of Astronomy and Space Science, Nanjing University, 163 Xianlin Avenue, Nanjing 210023, China}
\affil{Key Laboratory of Modern Astronomy and Astrophysics (Nanjing University), Ministry of Education, Nanjing 210023, China}

\author{Ray Furuya}
\affil{Institute of Liberal Arts and Sciences, Tokushima University, Minami Jousanajima-machi 1-1, Tokushima 770-850, Japan}

\author{David Berry}
\affil{East Asian Observatory, 660 N. A`oh\={o}k\={u} Place, University Park, Hilo, Hawaii 96720, USA}

\author{the JCMT BISTRO Survey Team}



\begin{abstract}

  We present the first high-resolution, submillimeter-wavelength polarimetric observations of -- and thus direct observations of the magnetic field morphology within -- the dense gas of the Pillars of Creation in M16.  These 850\,$\upmu$m observations, taken as part of the BISTRO (B-Fields in Star-forming Region Observations) Survey using the POL-2 polarimeter on the SCUBA-2 camera on the James Clerk Maxwell Telescope (JCMT), show that the magnetic field runs along the length of the pillars, perpendicular to, and decoupled from, the field in the surrounding photoionized cloud.  Using the Chandrasekhar-Fermi method we estimate a plane-of-sky magnetic field strength of $170-320\,\upmu$G in the Pillars, consistent with their having been formed through compression of gas with initially weak magnetization.  The observed  magnetic field strength and morphology suggests that the magnetic field may be slowing the pillars' evolution into cometary globules.  We thus hypothesize that the evolution and lifetime of the Pillars may be strongly influenced by the strength of the coupling of their magnetic field to that of their parent photoionized cloud -- i.e. that the Pillars' longevity results from magnetic support.

\end{abstract}



\section{Introduction} \label{sec:introduction}

One of the most iconic images taken by the Hubble Space Telescope (HST) was of the `Pillars of Creation' in M16 \citep{hester1996}.  These photoionized columns are typical of those found in high-mass star-forming regions throughout the interstellar medium.  M16 is a relatively local ($1.8\pm0.1$\,kpc; \citealt{dufton2006}), well-resolved, site of active ongoing star formation \citep{oliveira2008}, typical of regions forming high-mass ($>8$\,M$_{\odot}$) stars \citep{zinnecker2007}. We present the first detailed measurements of the magnetic field (hereafter B-field) in the densest parts of the pillars, taken as part of the B-Fields in Star-Forming Region Observations (BISTRO) survey \citep{wardthompson2017} on the James Clerk Maxwell Telescope (JCMT) using the Submillimeter Common-User Bolometer Array 2 (SCUBA-2) camera and its polarimeter POL-2.

Young massive stars produce sufficient high-energy photons to ionize a volume of their parent molecular cloud, thereby driving a shock into the cloud \citep{stromgren1939,zinnecker2007}.  These photoionized regions indicate ongoing high-mass star formation.  Complex structures can form in the dense gas at the shock interfaces \citep{spitzer1954} -- particularly, dense, neutral columns are frequently seen protruding into photoionized regions, most famously in M16.  The formation and evolution of these pillars remain disputed (\citealt{white1999}; \citealt{williams2001}; \citealt{ryutov2005}, hereafter Wh99; Wi01; R05 respectively), with the role of the B-field neither observationally nor theoretically well-constrained (\citealt{williams2007}; hereafter Wi07).  Near-infrared extinction observations of M16 suggest a difference in B-field direction between the Pillars and the surrounding photoionized cloud \citep{sugitani2007}, but cannot probe the dense gas of the Pillars themselves. 

The heads of the Pillars are dense, star-forming, molecular condensations (Wh99) interacting with the shock front associated with the young ($\sim1.3$ Myr; \citealt{bonatto2006}) high-mass cluster NGC6611 \citep{hillenbrand1993}.  Whether these condensations predate, or were formed by, the shock interaction is uncertain (Wh99; Wi01).  The heads are being destroyed by the interaction with NGC6611, with a lifetime of $\lesssim 3\times 10^{6}$\,yr \citep{mcleod2015}, and are thus likely to be considerably longer-lived than the lower-density pillars, whose estimated lifetime is a few $\times 10^5$\,yr (Wi01), suggesting that they will become disconnected cometary globules \citep{bertoldi1990}, unless another mechanism, such as a B-field, is at work. 

 We observed the Pillars of Creation in 850\,$\upmu$m polarized light with the POL-2 polarimeter (\citealt{friberg2016}) on the SCUBA-2 camera \citep{holland2013}, giving a map of the B-field in the dense gas of photoionized pillars unprecedented in sensitivity, area and resolution.  We observed Pillars I, II and III \citep{hester1996} at high S/N, and Pillar IV, the Spire and SFO30 (not shown) at lower S/N.

\section{Observations} \label{sec:observations}

The Eagle Nebula was observed in 20 separate 40-minute exposures between 6 June 2017 and 27 July 2017, with a total integration time of 14 hours.  The observations were taken in JCMT Band 2 weather, with atmospheric optical depth at 225 GHz, $\tau_{225}$, of $0.05<\tau_{225}<0.08$.  The BISTRO survey's observing strategy is described by \citet{wardthompson2017}.

The 850\,$\upmu$m POL-2 data were reduced using the $pol2map$ routine\footnote{http://starlink.eao.hawaii.edu/docs/sc22.pdf}, recently added to \textsc{smurf} \citep{berry2005,chapin2013}.  The reduction process is described in detail by \citet{kwon2018}. The output Stokes $Q$, $U$ and $I$ maps are gridded to 4$^{\prime\prime}$ pixels and are calibrated in mJy/beam. The output vectors are debiased using the mean of their $Q$ and $U$ variances to remove statistical biasing in regions of low signal-to-noise.

Our final map has a FWHM resolution of 14.1$^{\prime\prime}$ (0.12\,pc; $\sim25\,000$\,AU), a diameter of 12$^{\prime}$ and an RMS noise level of  0.9\,mJy/beam in Stokes $Q$ and $U$ intensity on 14.1$^{\prime\prime}$ pixels.

\section{Results} \label{sec:results}

\begin{figure}
  \centering
  \includegraphics[width=0.8\textwidth]{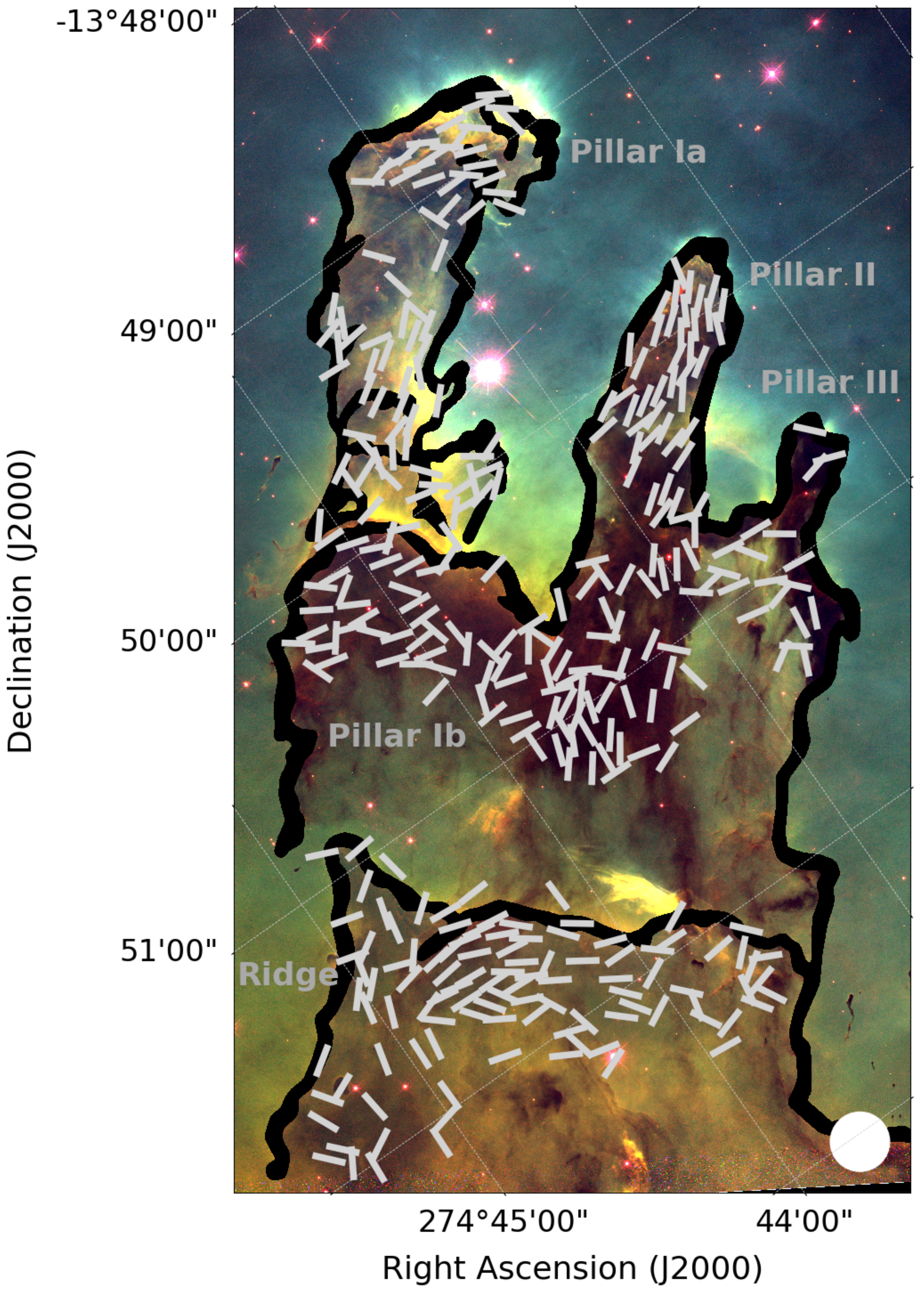}
  \caption{An illustrative figure of the BISTRO B-field vectors observed in the Pillars of Creation, overlaid on a HST 502\,nm, 657\,nm and 673\,nm composite \citep{hester1996}.  Vectors are gridded to 4$^{\prime\prime}$ (note oversampling), and have polarized intensity SNR $PI/\delta PI>2$.  Polarization angles are rotated by 90$^{\circ}$ to show B-field direction.  Vector length scale is arbitrary.  Black lines delineate the pillars.  Beam size is shown in lower right-hand corner.}
  \label{fig:HST}
\end{figure}

Figure~\ref{fig:HST} shows the observed B-field morphology in the Pillars.  We detect Pillars I, II and the material between their bases (the `Ridge') in polarized light, and marginally detect Pillar III.  The B-field clearly runs along the length of the pillars, apparently turning at the tips of the pillars (best seen in the head of Pillar I).  `Pillar I' has two separate components: Pillar Ia (north-west), located further along the line of sight than II and III, behind the source of ionizing photons; and  Pillar Ib (south-east), approximately equidistant with II and III \citep{pound1998,mcleod2015}.  The apparent change in field direction seen between Pillars Ia and Ib represents different field directions in the two pillars.

\begin{figure}
  \centering
  \includegraphics[width=0.7\textwidth]{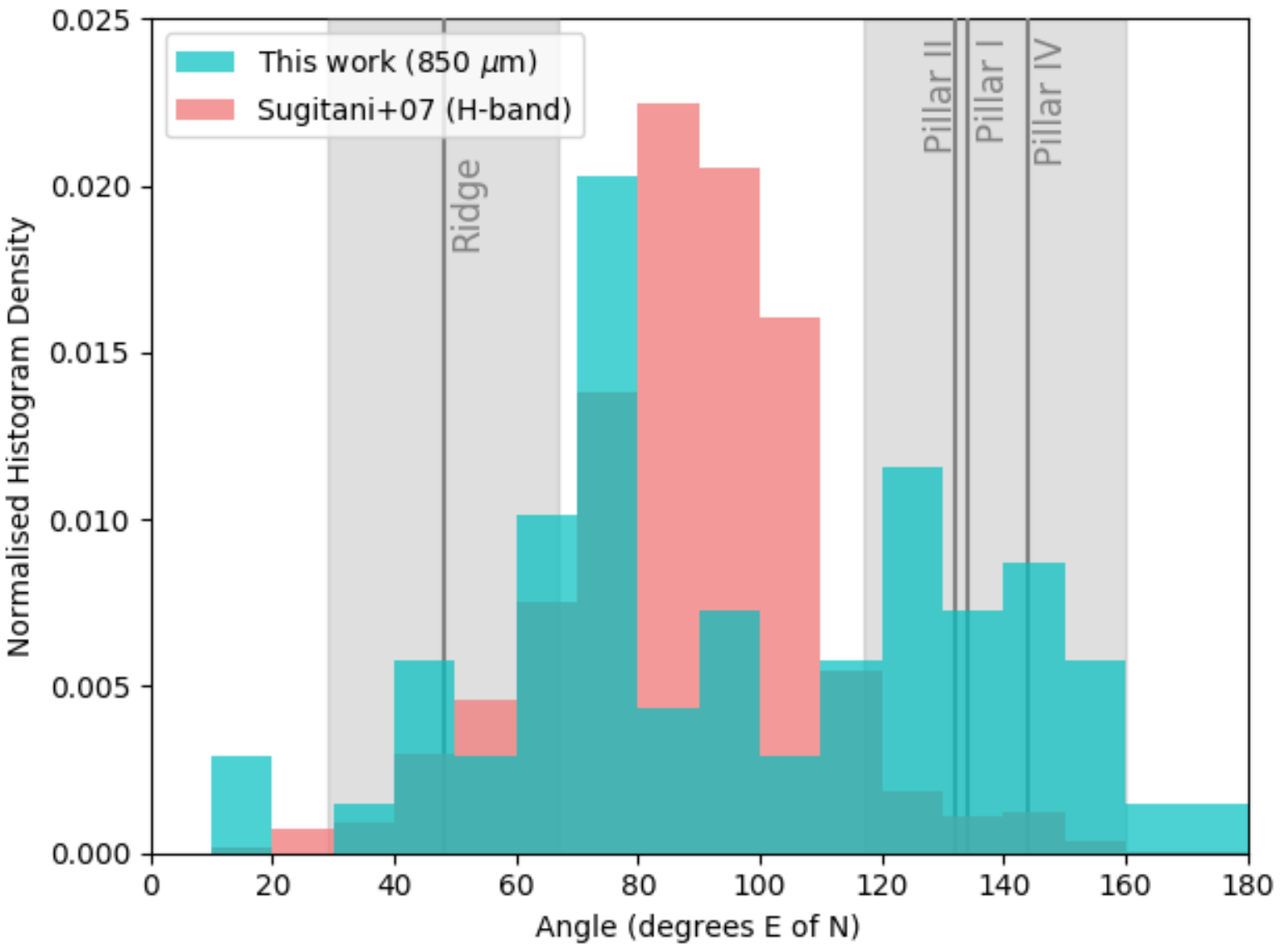}
  \caption{The distribution of the B-field vectors in the dense gas (blue -- this work; 850\,$\upmu$m dust polarization, 14.1$^{\prime\prime}$ pixels, $P/\delta P>3$, $I/\delta I>10$, $I>50$\,mJy\,beam$^{-1}$) and in the photoionized region (red -- H-band extinction polarization; \citealt{sugitani2007}).  Grey lines and shaded areas show the approximate orientations of Pillars I, II and IV and the Ridge, with the range derived from the Pillars' plane-of-sky aspect ratios.  Note how the red histogram peaks around $\sim90^{\circ}$ and the blue histogram peaks either side (roughly parallel to the Pillars and the Ridge, respectively).}
  \label{fig:histo}
\end{figure}

\begin{figure}
  \centering
  \includegraphics[width=0.45\textwidth]{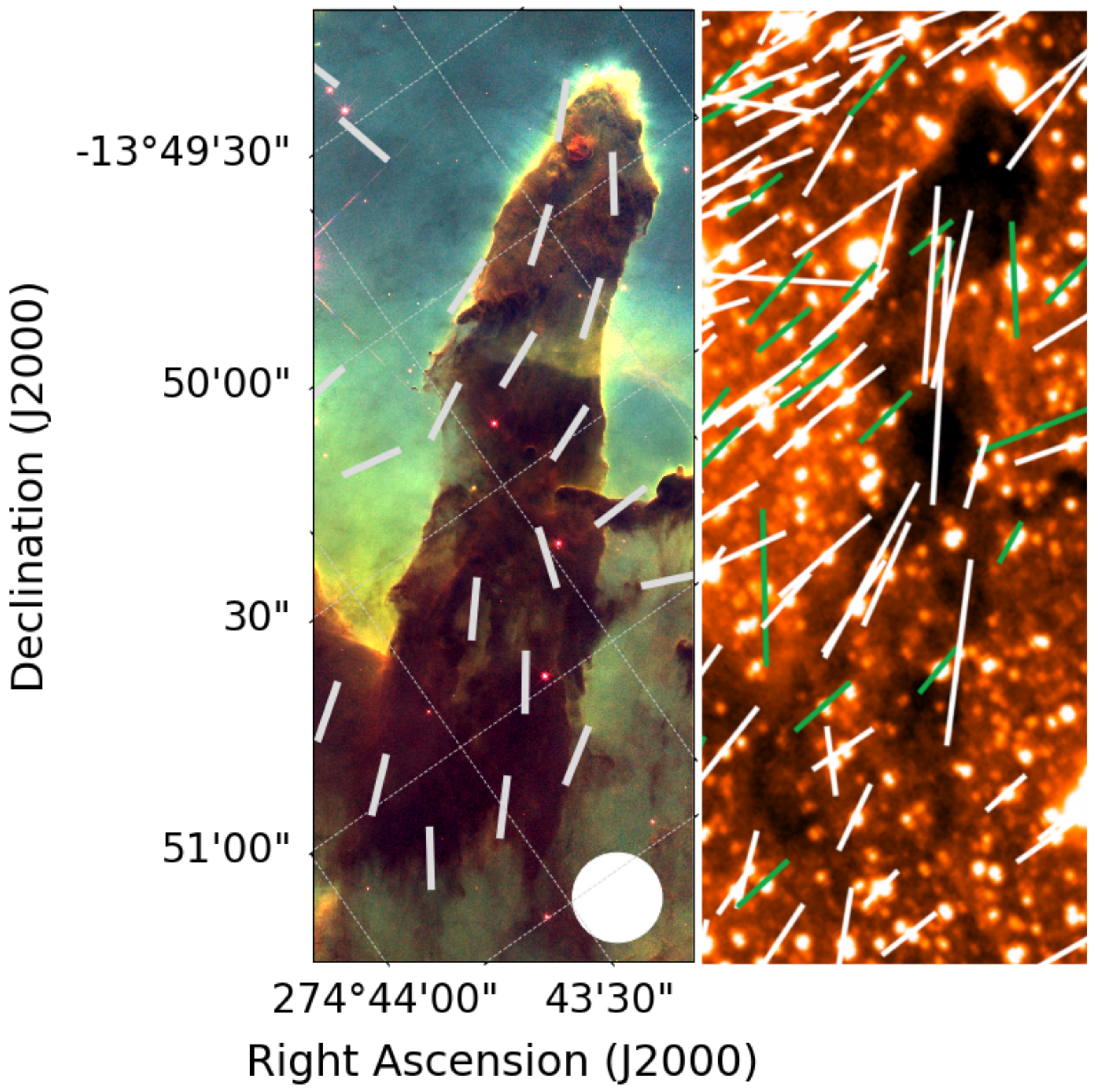}
  \caption{BISTRO B-field vectors overlaid on HST composite image of Pillar II, alongside H-band extinction polarimetry observations by \citet{sugitani2007}; excerpt from their Fig. 6 [permission obtained].  850$\upmu$m vectors (this work) have $P/\delta P>2$ and $I/\delta I>10$.  HST composite as in Figure~\ref{fig:HST}.  The B-field runs roughly parallel to the Pillar's axis.  No polarization is detected at the Pillar's tip -- this depolarization is consistent with a horseshoe-shaped B-field morphology on scales smaller than the beam.}
  \label{fig:zoom}
\end{figure}

The B-field geometry in the pillars is significantly different to that in the surrounding photoionized region, as measured using near-infrared extinction polarimetry \citep{sugitani2007}, as shown in Figure~\ref{fig:histo}.  The near-infrared vectors vary smoothly across the photoionized region, producing a singly-peaked distribution (at $\sim90^{\circ}$ east of north).  The B-field in the dense gas shows more complex behaviour, with field lines running roughly parallel to the length of the pillars.  The B-field distribution in the dense gas is bimodal, peaking at $\sim70^{\circ}$ (head of Ia, Ib, base of IV, Ridge) and $\sim140^{\circ}$ (length of Ia, II, IV), compared to mean pillar directions of $134^{\circ}\pm17^{\circ}$ in I, $132^{\circ}\pm12^{\circ}$ in II, $144^{\circ}\pm16^{\circ}$ in IV, and $48^{\circ}\pm19^{\circ}$ in the Ridge.  The B-field vectors observed in Pillar II -- upon which our subsequent analysis focusses -- are shown in detail in Figure~\ref{fig:zoom}.  The near-infrared polarization vectors observed by \citet{sugitani2007} in the vicinity of Pillar II are shown alongside.

\subsection{B-field strength}

We estimated the plane-of-sky B-field strength in Pillar II -- the most well-defined pillar, with velocity-coherent structure and a linear plane-of-sky morphology -- using the Chandrasekhar-Fermi method \citep{chandrasekhar1953}.

The Chandrasekhar-Fermi (CF) method provides an estimate of the plane-of-sky B-field strength by assuming that the variation in B-field around the mean field direction represents distortion of the B-field lines by non-thermal motions in the gas.  The plane-of-sky field strength ($B_{pos}$) is given by
\begin{equation}
B_{pos} = Q\sqrt{4\pi\rho}\frac{\sigma_{v}}{\sigma_{\theta}} \approx 9.3\sqrt{n({\rm H}_{2})}\frac{\Delta v}{\sigma_{\theta}}\,\upmu{\rm G}
\end{equation}
where $\rho$ is the gas density, $\sigma_{v}$ is the non-thermal gas velocity dispersion, $\sigma_{\theta}$ is the standard deviation in polarization angle about the mean field direction, and $Q$ is a factor of order unity that accounts for variation in the field on scales smaller than the beam.  We take $Q=0.5$ throughout \citep{ostriker2001,crutcher2004}.  The second form of the expression takes number density of molecular hydrogen ($n({\rm H}_{2})$) to be in cm$^{-3}$, FWHM non-thermal gas velocity dispersion ($\Delta v = \sigma_{v}\sqrt{8{\rm ln}2}$) to be in km\,s$^{-1}$, and $\sigma_{\theta}$ to be in degrees \citep{crutcher2004}.

\begin{figure}
  \centering
  \includegraphics[width=0.6\textwidth]{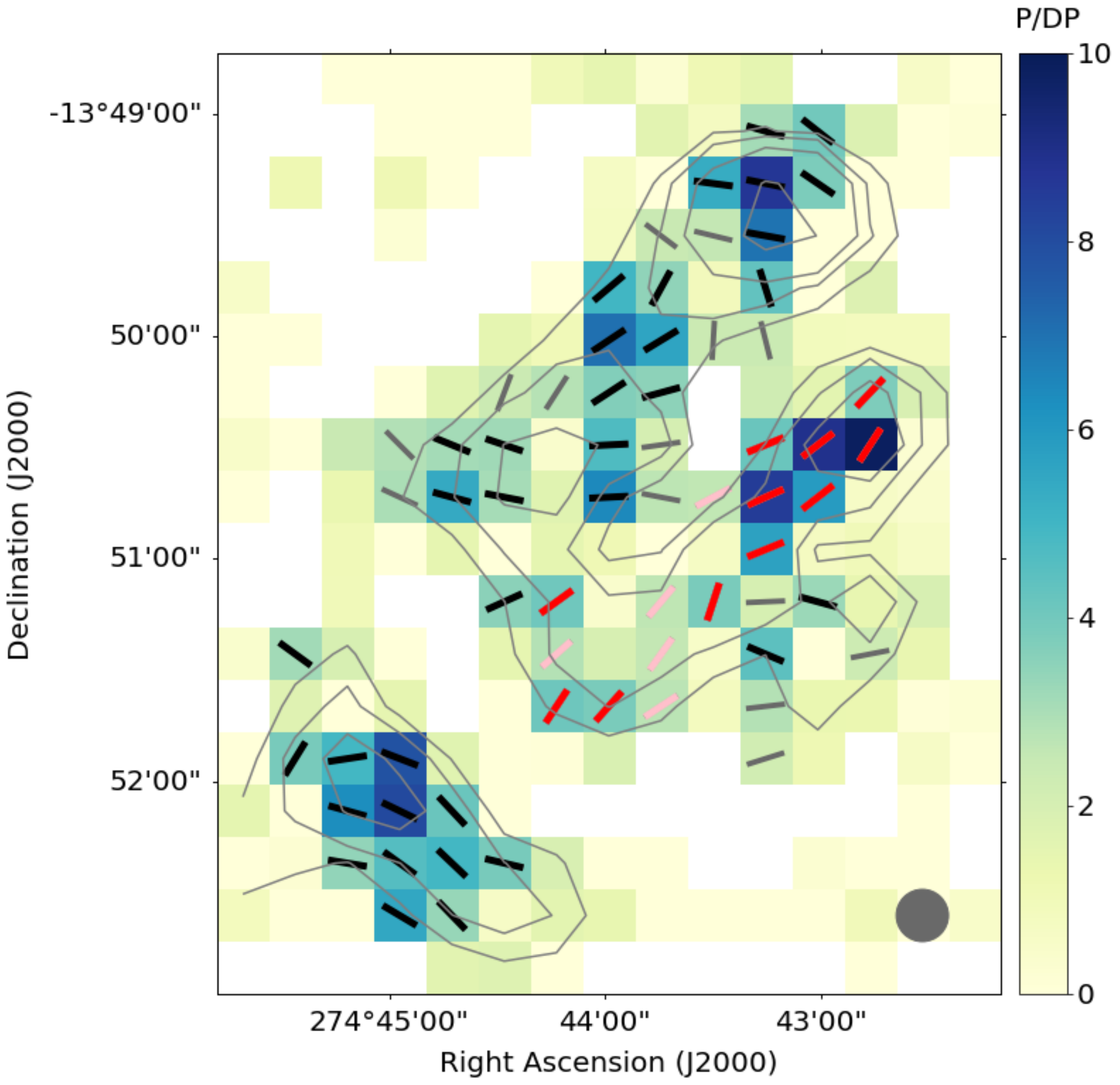}
  \caption{Signal-to-noise in $P/\delta P$, on statistically-independent pixels.  Red/pink vectors show pixels included in the CF analysis; black/grey vectors show pixels not included.  Pink/grey vectors have $3>P/\delta P>2$; red/black vectors have $P/\delta P>3$; all vectors have $I/\delta I>10$.  Contours show Stokes $I$ values of 50, 100, 200, 500 mJy\,beam$^{-1}$.  Beam size is shown in lower right-hand corner.}
  \label{fig:pdp}
\end{figure}

We binned the data to 14.1-arcsec resolution (statistically-independent pixels), and selected pixels with SNR in total intensity $I$ of $I/\delta I>10$ associated with Pillar II.  Of these, 16 have SNR in polarization fraction $P$ of $P/\delta P>2$, and 11 have $P/\delta P> 3$.  In order to mitigate against small sample size effects potentially introduced by using only the $P/\delta P> 3$ sample, we found the weighted standard deviations of both samples.  The $P/\delta P>3$ sample has a weighted dispersion in angle of $\sigma_{\theta}=14.4^{\circ}$, while the $P/\delta P>2$ sample has a very similar $\sigma_{\theta}=14.1^{\circ}$.  We thus adopt $\sigma_{\theta}\sim14.4^{\circ}$ as being a representative value.   We assume that all dispersion in the position angles of the vectors associated with the pillar represents dispersion about a uniform mean field direction.  As the measured angular dispersion is greater than the uncertainty on angle in our vectors, it is not necessary to correct the angular dispersion for measurement uncertainty \citep{pattle2017a}.  The $P/\delta P$ values of our data for 14.1-arcsec pixels are shown in Figure~\ref{fig:pdp}.

We took the gas density in the pillar to be $n({\rm H}_{2})=5\times 10^{4}$\,cm$^{-3}$ (R05), and the FWHM gas velocity dispersion to be in the range $\Delta v = 1.2 - 2.2$\,km\,s$^{-1}$, as measured by Wh99 in various dense gas tracers.  These linewidths are highly supersonic (Wh99), and so the correction for the thermal component is negligible.

We thus estimated a plane-of-sky B-field strength of $\sim170-320\,\upmu$G in Pillar II.  This value is intermediate between the B-field strengths of $\sim 10\,\upmu$G observed in relatively unperturbed gas in low-mass star-forming regions \citep{crutcher2012}, and of $\sim10^{3}\,\upmu$G observed in massive, gravitationally unstable structures in high-mass star formation sites (e.g. \citealt{curran2007,hildebrand2009,pattle2017a}).

\section{Discussion} \label{sec:discussion}

Simulations of photoionized regions suggest that B-field orientation is largely unchanged by the free passage of a plane-parallel shock front \citep{henney2009}.  Hence, we assume that the B-field in the photoionized region is representative of the B-field direction in the unshocked gas -- approximately parallel to the shock front.  For a weak initial B-field, field lines are predicted to become aligned parallel to the pillar's length in the pillar itself, while remaining approximately perpendicular in the surrounding photoionized region (Wi07; \citealt{mackey2011}, hereafter ML11).  This prediction results from otherwise quite different scenarios of magnetized pillar formation. 

Wi07 finds that, in two dimensions, when a shock propagates into a dense medium ($10^{4}$\,cm$^{-3}$) in which a denser core ($10^{5}$\,cm$^{-3}$) is embedded, a pillar forms behind the core, and the \emph{weak, plane-parallel} B-field in the dense medium is compressed.  Thus, the B-field strength is enhanced by pillar formation, with the field `bowing' into the material behind the pillar.  The pillar has a density of a few $\times10^{4}$\,cm$^{-3}$, while the surrounding ionized material has a density $\sim10^{2}$\,cm$^{-3}$.  (The pillar head has higher density.) \citet{arthur2011} find similar behaviour in three-dimensional simulations of expanding H\textsc{ii} regions, although with lower resolution. 

ML11 find that when a shock impinges on a set of approximately co-linear dense globules embedded in a much lower-density medium ($200$\,cm$^{-3}$; c.f. \citealt{mackey2010}) threaded by a \emph{weak, plane-parallel} B-field, a pillar-like feature forms behind the globules due to radiation-driven implosion and the rocket effect \citep{oort1955}.  These effects orient the B-field along the length of the forming pillar on timescales $\sim100$\,kyr.

Wi07 and M11 agree that a \emph{strong} plane-parallel initial B-field should deviate significantly from its initial orientation only in the pillar head (see also \citealt{henney2009}).  Our results do not match this scenario, strongly suggesting that the B-field in M16 was dynamically unimportant in the formation of the Pillars.

ML11 predict a B-field strength in the material \textit{around} the pillars of $<50\,\upmu$G, but do not quantitatively predict the B-field strength inside the pillars.  Our plane-of-sky B-field is in the ML11 `strong-field' regime, which they exclude for M16.  It is not clear how the gas compression necessary to increase the B-field could occur in this model.  (\citet{henney2009} predict volume-averaged B-field strengths to remain approximately constant with time within pillars formed behind individual globules.)  ML11 also show a B-field which while broadly orientated parallel to the pillar's length shows considerable disorder, whereas our observations show an ordered (albeit not well-resolved) B-field along the length of the Pillars.

\begin{figure}
  \centering
  \includegraphics[width=\textwidth]{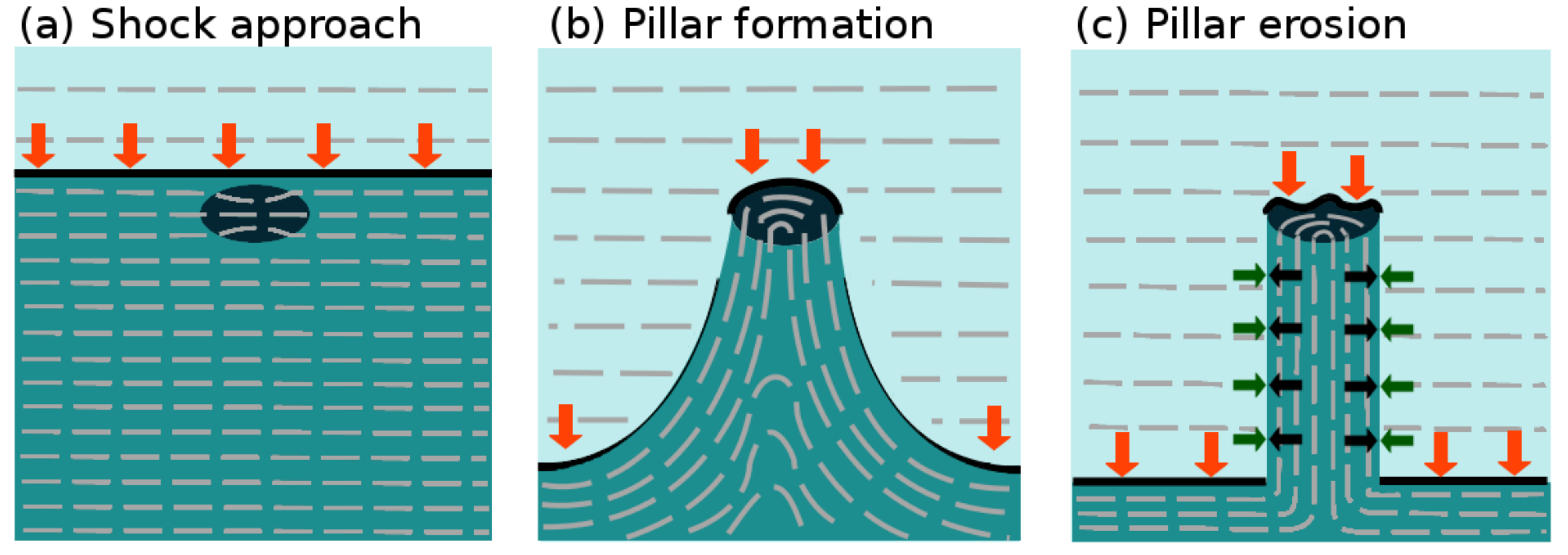}
  \caption{Our proposed evolutionary scenario: (a) an ionization front moving perpendicular to the ambient B-field approaches an existing over-density in the molecular gas.  (b) The ionization front is slowed by the over-density.  The flux-frozen B-field `bows' into the forming pillar. (c) The compressed B-field supports the pillar against radial collapse, but cannot support against longitudinal erosion by the shock interaction.  Dark blue represents molecular gas; light blue represents ionized material; black line indicates the shock front.  Grey dashed lines indicate local B-field direction.  Red arrows represent photon flux/ablation pressure, black arrows represent magnetic and internal gas pressure, and green arrows represent confining gas pressure, possibly supplemented by ram pressure.}
  \label{fig:cartoon}
\end{figure}

The Pillars are anchored to a larger cloud \citep{hester1996}, similar to the Wi07 scenario.  Moreover, the Wi07 simulations show the B-field compression necessary to significantly strengthen an initially dynamically unimportant field, albeit qualitatively and two-dimensionally.  We thus consider the Wi07 scenario to be broadly more consistent with our observations, and so illustrate it in Figure~\ref{fig:cartoon}.  However, both mechanisms could be involved in creating the observed B-field, and neither model quantitatively predicts B-field strength inside the pillars.  Detailed, three-dimensional, quantitative modelling of the B-field inside photoionized columns is needed to fully distinguish between these mechanisms.

\subsection{Pressure balance}

Magnetic pressure is given by $P_{B}=B^{2}/8\pi$.  Our measured plane-of-sky B-field, $170-320$\,$\upmu$G, implies $P_{B}/k_{\textsc{b}}\sim(0.9 - 3.0)\times10^{7}$\,K\,cm$^{-3}$.  R05 give an ablation pressure on the heads of the pillars of $1.6\times 10^{8}$\,K\,cm$^{-3}$, an order of magnitude higher than our inferred $P_{B}$.  This suggests that the B-field cannot support the pillars against longitudinal erosion by the shock front, unless the field is compressed in the pillar heads (which are not resolved by our observations).

The effective gas pressure within the Pillars is $P_{g,int} = nk_{\textsc{b}}T_{eff}$, where $T_{eff}$ is the effective gas temperature and $n$ is number density of particles.  Taking $n\approx n({\rm H}_{2})=5\times10^{4}$\,cm$^{-3}$ and $T=20$\,K (Wh99), $P_{g,int}/k_{\textsc{b}}=1.0\times10^{6}$\,K\,cm$^{-3}$, an order of magnitude lower than our $P_{B}$.  However, Wh99 and Wi01 argue that non-thermal gas motions create an effectively hydrostatic pressure within the Pillars.  The Wh99 FWHM gas velocity dispersion range $\Delta v=c_{eff}\sqrt{8\ln2}=1.2-2.2$\,km\,s$^{-1}$ thus represents an effective sound speed $c_{s,eff} = (k_{\textsc{b}}T_{eff}/\mu m_{\textsc{h}})^{0.5} \approx 0.51 - 0.93$\,km\,s$^{-1}$ and so, for a mean molecular weight $\mu=2.8$, $P_{g,int}/k_{\textsc{b}}=(0.4-1.5)\times10^{7}$\,K\,cm$^{-3}$, very comparable to our inferred $P_{B}$.  (This result follows naturally from the assumptions of the CF analysis.)

\citet{hester1996} argue that atomic hydrogen number density $n({\rm H})\sim29$\,cm$^{-3}$ in the M16 photoionized region.  Simulations take $n({\rm H})\sim10^{2}$\,cm$^{-3}$ (assumed by Wi01 and ML11; predicted by Wi07).  Using $n\approx2n({\rm H})=58$\,cm$^{-3}$ (assuming $n_{e}\approx n({\rm H})$ and that the number fraction of helium atoms is small), and taking $T=8000$\,K \citep{hester1996, garcia-rojas2006}, this implies an external gas pressure $P_{g,ext}/k_{\textsc{b}}\sim 4.6\times10^{5}$\,K\,cm$^{-3}$ on the Pillars.  Using $n\approx2n(\rm H)=400$\,cm$^{-3}$, $P_{g,ext}/k_{\textsc{b}}\sim 3.2\times 10^{6}$\,K\,cm$^{-3}$, still an order of magnitude lower than our $P_{B}$ and $P_{g,int}$ values.

\citet{higgs1979} find a non-thermal velocity dispersion in the photoionized gas of M16 of $\sigma_{v}=11.5$\,km\,s$^{-1}$.  If these non-thermal motions create a hydrostatic pressure on the Pillars, then $c_{s,eff}=\sqrt{c_{s}^{2}+\sigma_{v}^{2}}\approx14.1$\,km\,s$^{-1}$ in the photoionized region, equivalent to $T_{eff}\approx 3.4\times10^{4}$\,K if $\mu = 1.4$ in the ionized material (consistent with the $\mu$ value we use in the molecular gas).  For $n\approx2n({\rm H})=400$\,cm$^{-3}$, this implies $P_{g,ext}/k_{\textsc{b}}\sim1.4\times10^{7}$\,K\,cm$^{-3}$, comparable to our inferred internal $P_{B}$.

The above analysis assumes a uniform-density, i.e. non-self-gravitating, pillar.  Pillar II has radius $\sim0.15$\,pc, and so line mass $M/L=\mu m_{\textsc{h}}n\pi r^{2}\approx250$\,M$_{\odot}$\,pc$^{-1}$, assuming cylindrical symmetry (taking $\mu = 2.8$ and $n=n({\rm H}_{2})=5\times10^{4}$\,cm$^{-3}$).  If non-thermal gas motions within the pillars create hydrostatic pressure, the critical line mass (\citealt{stodolkiewicz1963}; \citealt{ostriker1964}) is
\begin{equation}
 \left(\frac{M}{L}\right)_{crit} = \frac{2 c_{eff}^{2}}{G}.
\end{equation}
For $c_{eff}\approx0.51-0.93$\,km\,s$^{-1}$, $(M/L)_{crit}\approx120-400$\,M$_{\odot}$\,pc$^{-1}$, comparable to the observed $M/L$. Thus, there may be some concentration of mass towards its axis, somewhat lowering $P_{g,int}$ at the H\textsc{ii} region boundary.  However, the B-field will provide significant support against radial gravitational collapse, with the observed B-field geometry resisting radial motion of material.

We note that these estimates are accurate only to order of magnitude.  Our results broadly suggest that the pillar walls are in approximate pressure equilibrium, with $P_{B}$ and $P_{g,int}$ supporting against $P_{g,ext}$, and also that, contrary to common assumptions, the pillar's self-gravity is non-negligible.  Both $P_{g,int}$ and $P_{g,ext}$ require a non-thermal component in order to be comparable to our inferred $P_{B}$.  Other sources of external pressure could include ram pressure due to flow of material across the ionization front into the pillar (e.g. \citealt{henney2009}).

\subsection{The Alfv\'{e}n velocity}

Our favoured scenario requires (a) the flux-frozen (infinite conductivity) approximation \citep{alfven1942, crutcher2012} to hold (neutral and ionized material are collisionally coupled; flow across field lines is forbidden), and (b) the pillars to form faster than the compressed B-field can relax to a lower-energy configuration, i.e. the photoionized region must expand faster than the Alfv\'{e}n velocity ($v_{\textsc{a}}$; \citealt{alfven1942}).  The photoionized region is expanding at a rate of $\sim2-10$\,km\,s$^{-1}$ (Wi01; \citealt{mcleod2015}).  For a representative B-field strength of $250\,\upmu$G and $n=5\times10^{4}$\,cm$^{-3}$ in the pillars, $v_{\textsc{a}}=B/\sqrt{\mu_{0}\mu m_{\textsc{h}}n}\sim 1.5$\,km\,s$^{-1}$.  Since $B\propto\sqrt{n}$ in flux-frozen plasma, this value should apply throughout the pillars' lifetimes, suggesting that  they could have formed too quickly for the B-field to react, allowing the observed, highly pinched, geometry to form (see Figure~\ref{fig:cartoon}).

The (flux-frozen) B-field geometry should allow longitudinal motion of material along the pillars, but strongly resist motion across the pillars that would lead to radial collapse.  This suggests that the predicted evolution of the pillars into disconnected cometary globules \citep{bertoldi1990} may be considerably slowed by the effects of the B-field geometry.

\section{Summary} \label{sec:summary}

We have observed the dense gas of the Pillars of Creation in M16 in 850$\upmu$m polarized light using the POL-2 polarimeter on the JCMT.  We find that the B-field in the Pillars is ordered, running along the length of the pillars, with a plane-of-sky field strength of $\sim 170-320\,\upmu$G, estimated using the Chandrasekhar-Fermi method.  The observed morphology is consistent with the field being dynamically negligible in the Pillars' formation.  However, the current B-field strength suggests that magnetic pressure provides significant support against both gravitational and pressure-driven radial collapse of the pillars, and may be slowing the Pillars' evolution into cometary globules.  We hypothesize that the persistence of such photoionized columns as objects connected to their parent molecular cloud may be related to the geometry of their B-fields, and specifically to the relative orientation of the B-fields in the pillars and their surrounding photoionized regions.  The BISTRO project is currently surveying B-fields in the dense gas of many nearby high-mass star-forming regions, thus allowing further testing of this hypothesis in the immediate future. 

\acknowledgments
The JCMT is operated by the East Asian Observatory on behalf of the National Astronomical Observatory of Japan, the Academia Sinica Institute of Astronomy and Astrophysics, the Korea Astronomy and Space Science Institute, the National Astronomical Observatories of China and the Chinese Academy of Sciences (Grant No. XDB09000000), with additional funding support from the Science and Technology Facilities Council (STFC) of the United Kingdom (UK) and participating universities in the UK and Canada.  The JCMT was historically operated by the Joint Astronomy Centre on behalf of the STFC of the UK, the National Research Council of Canada and the Netherlands Organisation for Scientific Research.  Additional funds for SCUBA-2 and POL-2 were provided by the Canada Foundation for Innovation.  The data used herein have project code M17BL011.  KP \& DWT acknowledge the STFC (Grant No. ST/M000877/1); KP \& SPL, the Ministry of Science and Technology (Taiwan) (Grant No. 106-2119-M-007-021-MY3); WK, the Basic Science Research Program through the National Research Foundation of Korea (NRF-2016R1C1B2013642).  This research used: the Canadian Advanced Network for Astronomy Research, the Canadian Astronomy Data Centre, the NASA Astrophysics Data System.  The authors recognize and acknowledge the very significant cultural role, and reverence, of the summit of Maunakea within the indigenous Hawaiian community. 

%

\vspace{5mm}
\facilities{James Clerk Maxwell Telescope (JCMT)}

\software{Starlink software \citep{currie2014},
  astropy \citep{astropy}
}

\end{document}